\documentclass[submission,copyright,creativecommons]{eptcs}
\providecommand{\event}{FROM 2019} 
\usepackage{breakurl}             
\usepackage{underscore}           
\usepackage{alltt}
\usepackage{proof}
\usepackage{amssymb}
\usepackage{amsthm, amsmath}
\usepackage{stackengine}

\usepackage{enumitem}
\setlist{nolistsep}

\newtheorem{theorem}{Theorem}[section]

\newtheorem{lemma}[theorem]{Theorem}

\def\eqbydef{\mathrel{\ensurestackMath{\stackon[1pt]{=}{\scriptstyle\Delta}}}}

\title{A Formal Semantics of Findel in Coq (Short Paper)}
\author{Andrei Arusoaie
\institute{Faculty of Computer Science - UAIC, Ia{\c s}i, Romania}
\email{andrei.arusoaie@uaic.ro}}
\def\authorrunning{Andrei Arusoaie}
\def\titlerunning{A Formal Semantics of Findel in Coq}
\begin{document}
\maketitle

\begin{abstract}
We present the first formal semantics of Findel - a DSL for specifying financial derivatives. 
The semantics is encoded in Coq, and we use it to prove properties of several Findel contracts.
\end{abstract}

\section{Introduction}

Financial derivatives are contracts between two or more parties whose value is based on underlying financial assets (e.g., bonds, comodities, currencies, etc.).  Typically, these contracts are expressed using natural language on a written document which is authorized by a trusted authority. However, natural language is often ambiguous and lacks precision. To overcome this issue, domain specific languages (DSLs) for expressing precise financial agreements have been developed~\cite{Christiansen2013AnAP, Jones00composingcontracts}.
Findel~\cite{findel} is yet another DSL in this category, which is restricted to financial derivatives. Findel is designed to be executed on the Ethereum blockchain platform and thus, it 
enables the use of smart contracts to facilitate derivatives trading. In~\cite{findel}, the language is advertised as being capable of ``expressing the most common derivatives''.

A fixed rate currency exchange~\cite{findel} contract, can be expressed in Findel as follows:
\begin{center}
\begin{minipage}{.8\textwidth}
$\mathit{FRCE} \eqbydef$ {\tt
And(Give(Scale(11, One(USD))), Scale(10, One(EUR)))}
\end{minipage}
\end{center}
\noindent
Each Findel contract has two parties: an \emph{issuer} and an \emph{owner}. Let us assume that {\tt Alice} is the issuer. If {\tt Bob} joins the contract, then  {\tt Bob} becomes an owner and the contract is now ready for execution. {\tt And}, {\tt Give},  {\tt Scale}, and {\tt One} are called \emph{primitives}, i.e., the building blocks of Findel contracts. {\tt And} specifies that its both (sub)contracts {\tt Give(Scale(11, One(USD)))} and  {\tt Scale(10, One(EUR))} are executed sequentially. {\tt Give} swaps the issuer and the owner. {\tt Scale} multiplies the value of the contract by a specified factor. Finally, {\tt One} is the only primitive that actually performs a transfer from the issuer to the owner. So, $\mathit{FRCE}$ specifies that  {\tt Bob} gives to {\tt Alice} 11 dolars \underline{and} receives 10 euros from {\tt Alice}.

Findel has an implementation in Solidity, which is available on Github: \url{https://github.com/cryptolu/findel}. Findel contracts can be registered inside an Ethereum smart contract and then executed. Once deployed, a contract cannot be changed and thus, it very important for someone who deploys a contract to make sure that the terms of the involved transactions cannot be violated. For example, it is important for {\tt Alice} to know that once {\tt Bob} joins the contract she receives 11 dollars. Conversely, it is important for {\tt Bob} to know that {\tt Alice} will pay him back 10 euros precisely as specified by the contract. 

Ideally, a trustworthy organization (e.g., a bank) may want to issue Findel contracts together with proof certificates to ensure that both the issuer and the owner have certain rights and obligations.

\emph{Contributions.}
We take the challenge to prove and certify properties about Findel contracts. 
To achieve that, we formally define the semantics of Findel primitives in Coq.
We add to our semantics several components in order to simulate an execution environment, where multiple contracts can be executed, events are triggered, and transactions are registered in a ledger. 
This environment, hereafter called \emph{marketplace}, enables us to prove not only properties of Findel primitives, but also properties that depend on other factors (external data providers, time, events). 
Besides simple examples, we approach several non-trivial contracts and we highlight problems that contracts may hide. Note that we formalise the existing semantics of Findel as presented in~\cite{findel}, and we do not improve or modify the language.


\begin{table}
\small
\centering
\begin{tabular}{| l | l | }
\hline
\bf Primitive & \bf Informal semantics\\
\hline\hline
{\tt Zero} & Do nothing.\\
\hline

{\tt One(}\textit{currency}{\tt )} & Transfer 1 unit of {\textit{currency}} from issuer to the owner.\\
\hline
\hline
{\tt Scale(}$k$, $c${\tt )} & Multiply all payments of $c$ by a constant factor of $k$.\\
\hline
{\tt ScaleObs(}$a$, $c${\tt )} & Multiply all payments of $c$ by a factor obtained from address $a$.\\
\hline
{\tt Give(}$c${\tt )} & Swap parties of $c$.\\
\hline
{\tt And(}$c_1$, $c_2${\tt )} & Execute $c_1$ and then $c_2$.\\
\hline
{\tt Or(}$c_1$, $c_2${\tt )} & Give the owner the right to execute either $c_1$ or $c_2$ (not both).\\
\hline
{\tt If(}$a$, $c_1$, $c_2${\tt )} & If $b$ is true, execute $c_1$, else execute $c_2$, where $b$ is obtained from address $a$.\\
\hline
{\tt Timebound(}$t_0$, $t_1$, $c${\tt )} & Execute $c$, if the current timestamp is within $[t_0, t_1]$.\\
\hline
\end{tabular}
\caption{The informal semantics of Findel primitives~\cite{findel}.}
\label{tbl:informal}
\end{table}

\section{The informal semantics of Findel}
\label{sec:informal}

A Findel contract is a tuple with three components: a \emph{description}, an issuer and an owner. A Findel description is essentially a tree with \emph{basic primitives} as leaves and \emph{composite primitives} as internal nodes. The list of available primitives and their informal semantics is shown in Table~\ref{tbl:informal}. {\tt Zero} and {\tt One} are basic primitives, while the others are composite. 
The execution model~\cite{findel} of a Findel contract is:

\begin{enumerate}
\item The first party \emph{issues} a contract and becomes its issuer. This is a mere declaration of the issuer's desire to conclude an agreement and entails no obligations.
\item The second party \emph{joins} the contract and becomes its owner. As a consequence, both parties accept the specified rights and obligations.
\item The contract is executed immediately:
	\begin{enumerate}
		\item Let the current node be the root node of the contract description.
				If the current node is either {\tt Or} or {\tt Timebound} with $t_0 > \mathit{now}$, postpone the execution: issue a new Findel contract, with the same parties and the current node as root. The owner can demand later its execution.
		\item Otherwise, execute all sub-nodes recursively.
		\item Delete the contract.
	\end{enumerate}
\end{enumerate}

A Findel contract is not guaranteed to be executed: only the owner can trigger the execution for issued contracts.
It is worth noting that the execution of a contract may produce new contracts (if {\tt Or} or {\tt Timebound} are present) or it can affect the balances of the parties. The current implementation of Findel does not enforce any constraints on the balances of the users that prevents them from building up debt.

\section{The formal semantics in Coq}
\label{sec:formal}

\subsection{Syntax}
\label{sec:syntax}
The syntax of Findel is fairly small in size and it is quite easy to encode it in Coq using {\tt Inductive} as shown in Figure~\ref{fig:syntax}.
\begin{figure}
\begin{center}
\begin{minipage}{.8\linewidth}
\small
\begin{alltt}
Inductive Primitive :=
(* basic primitives *)
| Zero      :                                      Primitive
| One       : Currency ->                          Primitive
(* composite primitives *)
| Scale     : nat -> Primitive ->                  Primitive 
| ScaleObs  : Address -> Primitive ->              Primitive
| Give      : Primitive ->                         Primitive
| And       : Primitive -> Primitive ->            Primitive
| Or        : Primitive -> Primitive ->            Primitive
| If        : Address -> Primitive -> Primitive -> Primitive
| Timebound : nat -> nat -> Primitive ->           Primitive.
\end{alltt}
\end{minipage}
\caption{The syntax of Findel in Coq.}
\label{fig:syntax}
\end{center}
\end{figure}
\noindent
For convenience, several additional primitives are introduced in~\cite{findel}: {\tt At}, {\tt Before}, {\tt After}, and {\tt Sell}. In Coq, these are simple definitions. We show here only the definition for {\tt At}:
\begin{center}
\small
{\tt Definition At (t : nat) (p : Primitive) := Timebound (t - $\Delta$) (t + $\Delta$) p.}
\end{center}

\noindent
$\Delta$ is a just parameter which is used to adjust intervals for accepting transactions.
Having these new constructs enables us to define interesting contracts in Coq. For instance, here is a zero-coupon-bond contract, where the issuer asks for 10 dollars and will pay an eventual owner 11 dollars after 1 year:

\begin{center}
\begin{minipage}{.95\textwidth}
\small
$\mathit{ZCB} \!\eqbydef\!$ {\tt (And (Give (Scale (One USD)  10)) (At (\texttt{now} + 1 \texttt{yr}) (Scale (One USD) 11)))
}
\end{minipage}
\end{center}

\subsection{Semantics}

In Coq, we provide a function called {\tt exec} which executes primitives recursively. The function takes as inputs the description id, a scale factor, the issuer, the owner, the current time, an external gateway, the ledger, and a fresh id. The base cases of {\tt exec} are: for {\tt Zero} it does nothing, for {\tt One} it modifies the balances of the participants and the ledger, for {\tt Or} and {\tt Timebound} it generates new contracts. 
The gateway is used to retrieve external data, and it is modeled as a map from addresses to values.
The function {\tt exec} returns a tuple: $\langle$ the updated balance, the generated contracts, a fresh identifier, the updated ledger $\rangle$. 
Contracts are denoted as $[i,d,p,I,O,p_O,s]$, where $i$ is the contract id, $d$ is the corresponding description, $I$ is the issuer, $O$ is the owner, $p_O$ is the the proposed owner, and $s$ is the scale of the contract. 

The marketplace (i.e., the environment where contracts are executed) is a tuple with several components $\langle \mathcal{C}, \mathcal{D}, \mathcal{B}, t, \mathcal{G}, i, \mathcal{L}, \mathtt{E} \rangle$ encoded as a record in Coq: 
$\mathcal{C}$ - the list of the issued contracts;
$\mathcal{D}$ - the list of the available descriptions;
$\mathcal{B}$ - the current balance;
$t$ - the current timestamp, i.e., a natural number;
$\mathcal{G}$ - the external gateway, i.e., a map from addresses to values, used to execute {\tt ScaleObs} and {\tt If};
$i$ - a fresh identifier i.e., a natural number which is always fresh;
$\mathcal{L}$ - a global ledger, i.e., the ordered list of the performed transactions;
$\mathtt{E}$ - the list of events, which are triggered during execution (e.g., {\tt Deleted}, {\tt Executed}, {\tt IssuedFor}).
We show below the semantical rules that specify how the marketplace evolves.\\

\noindent
{\bf [Issue]} When issued, a contract is added to the list of issued contracts in the marketplace having a unique id $i$. The owner field contains the address of the issuer, while the proposed owner field contains the address of the intended owner. 
The primitive of the contract is initialized from an existing description. Also an event {\tt IssuedFor} is triggered, and the global fresh id is incremented:

$$
\infer[]{
\langle [i,\mathit{id}(dsc),\mathit{primitive}(dsc),I,I,p_O,s] : \mathcal{C}, \mathcal{D}, \mathcal{B}, t, \mathcal{G}, i + 1, \mathcal{L}, (\mathtt{IssuedFor}~p_O~i) : \mathtt{E} \rangle
}{
\langle \mathcal{C}, dsc \in \mathcal{D}, \mathcal{B}, t, \mathcal{G}, i, \mathcal{L}, \mathtt{E} \rangle
}
$$
\noindent
{\bf [Join]} Joining a contract is the most complex operation and requires several conditions. First, $(A)$ $O = p_O$, i.e., the owner is the proposed owner. Second, the root node should not be an {\tt Or}: $(B)$ $p \neq \mathtt{(Or~\_~\_)}$. Third, the execution is limited within a time interval by its description $d$: $(C)$ $\mathit{start}(d) \leq \mathit{now} \leq \mathit{end}(d)$. Finally, the execution should be successful: $(D)$ $\mathtt{exec}(\mathit{p}, \mathit{c\_id}, \mathit{dsc\_id}, \mathit{s}, \mathit{I}, \mathit{O}, \mathit{t}, \mathcal{G}, \mathcal{L}, i) = \langle \mathcal{B'}, \mathcal{C'}, i', \mathcal{L'} \rangle$. The rule for joining a contract is shown below. Note that {\tt E} is enriched with an {\tt Executed} event.
$$
\infer[]{
\langle \mathcal{C}, \mathcal{D}, \mathcal{B'}, t, \mathcal{G'}, i', \mathcal{L'}, (\mathtt{Executed}~\mathit{c\_id}) : \mathtt{E} \rangle
}{
\langle \{ [\mathit{c\_id},\mathit{dsc\_id},\mathit{p},I,I,p_O,s] \} \cup \mathcal{C}, \mathcal{D}, \mathcal{B}, t, \mathcal{G}, i, \mathcal{L}, \mathtt{E} \rangle 
& (A) & (B) & (C) & (D)
}
$$
\noindent
{\bf [Join {\tt OR}]} If the primitive of the current contract is $({\mathtt Or}~c_1~c_2)$ then the owner can execute either $c_1$ or $c_2$: $(D')$ $\mathtt{exec}(\square, \mathit{c\_id}, \mathit{dsc\_id}, \mathit{s}, \mathit{I}, \mathit{O}, \mathit{t}, \mathcal{G}, \mathcal{L}, i) = \langle \mathcal{B'}, \mathcal{C'}, i', \mathcal{L'} \rangle$. Here, $\square$ is either $c_1$ or $c_2$: 
$$
\infer[]{
\langle \mathcal{C}, \mathcal{D}, \mathcal{B'}, t, \mathcal{G'}, i', \mathcal{L'}, (\mathtt{Executed}~\mathit{c\_id}) : \mathtt{E} \rangle
}{
\langle \{ [\mathit{c\_id},\mathit{dsc\_id},({\mathtt Or}~c_1~c_2),I,I,p_O,s] \} \cup \mathcal{C}, \mathcal{D}, \mathcal{B}, t, \mathcal{G}, i, \mathcal{L}, \mathtt{E} \rangle 
& (A) & (C) & (D')
}
$$
\noindent
{\bf [Fail]} If the execution of a contract fails (e.g., if contract is expired or the gateway does not provide fresh data), i.e., $(F)$ $\mathtt{exec}(\mathit{p}, \mathit{c\_id}, \mathit{dsc\_id}, \mathit{s}, \mathit{I}, \mathit{O}, \mathit{t}, \mathcal{G}, \mathcal{L}, i) = \bot$, then the {\tt Deleted} event is triggered:
$$
\infer[]{
\langle \mathcal{C}, \mathcal{D}, \mathcal{B}, t, \mathcal{G}, i, \mathcal{L}, (\mathtt{Deleted}~\mathit{c\_id}) : \mathtt{E} \rangle
}{
\langle \{ [\mathit{c\_id},\mathit{dsc\_id},\mathit{p},I,I,p_O,s] \} \cup \mathcal{C}, \mathcal{D}, \mathcal{B}, t, \mathcal{G}, i, \mathcal{L}, \mathtt{E} \rangle 
& (A) & (B) & (C) & (D)
}
$$
\noindent
{\bf [Tick]} Since the execution of the Findel contracts depends on the current timestamp, the marketplace contains a global time. The tick rule increases the global time $t$:
$
\vcenter{
\infer[]{
\langle \mathcal{C}, \mathcal{D}, \mathcal{B}, t + 1, \mathcal{G}, i, \mathcal{L}, \mathtt{E} \rangle
}{
\langle \mathcal{C}, \mathcal{D}, \mathcal{B}, t, \mathcal{G}, i, \mathcal{L}, \mathtt{E} \rangle
}}
$
\section{Proving properties of Findel contracts}
\label{sec:proofs}

A Findel contract specifies certain rights and obligations for parties. 
We show here examples of contracts and some interesting properties that we prove in Coq.
We will see that, although they are are small is size, contracts may hide unexpected behaviors.
Hereafter, we assume that {\tt Alice} is the issuer and {\tt Bob} is the owner. For each contract, we formalise the desired properties as theorems and we prove them in Coq.\\[.5ex]

\noindent
\textbf{I. Fixed-rate currency exchange.}
Recall the fixed rate currency exchange $FRCE$ contract, where {\tt Bob} gives {\tt Alice} 11 dollars \underline{and} receives 10 euros from {\tt Alice}. In Coq,  we are able to prove the next lemma:
\begin{lemma}
If $FRCE$ is executed, then {\tt Alice} receives 11 dollars.
\end{lemma}
\noindent
The lemma is straightforward: the hypothesis imply that the {\bf [Join]} rule applies successfully. In this case, a transaction from {\tt Bob} to {\tt Alice} involving an amount of 11 dollars is registered in the ledger.
Conversely, one may want to prove that {\tt Bob} also receives 10 euros from {\tt Alice}, which is also straightforward in Coq: 
\begin{lemma}
If $FRCE$ is executed, then {\tt Bob} receives 10 euros from {\tt Alice}.
\end{lemma}

\noindent
Note that the lemmas above hold, no matter what other contracts are executed in the marketplace. If one of the participants does not have enough money to pay then they can build up debt. The current semantics of Findel does not prevent from building up too much debt~\cite{findel} and we do not formalize that in Coq.\\[.5ex]

\noindent
\textbf{II. External gateway rate - currency exchange.}
A more interesting contract is a currency exchange that uses a rate retrieved from an external source (a gateway). Consider the following Findel description:

\begin{center}
\noindent
$\mathit{EXT} \!\eqbydef\!$
{\tt (And (Give\,(Scale n\,(One EUR))) (ScaleObs ADDR (Scale n (One USD))))\\
}
\end{center}

\noindent
$\mathit{EXT}$ specifies that the issuer is supposed to receive {\tt n} {\tt EUR} and will pay to the owner ({\tt rate * n}) {\tt USD}, where the {\tt rate} is obtained from an external address {\tt ADDR} using the gateway.
A non-obvious problem is that it is possible for the gateway to fail for various reasons (e.g., data expired, invalid address). In such cases, the contract execution fails and the specified agreement between parties fails as well. 
Even if the balance of the parties is not affected, an intruder can determine certain contracts to fail on purpose.

In~\cite{findel}, the owner is ``advised'' to update gateways ``shortly before execution''. Also, the gateway is assumed to be ``trustworthy''. Under the same assumptions we prove in Coq the next theorems:

\begin{lemma}
If $\mathit{EXT}$ is executed then {\tt Alice} receives an amount of {\tt n} euros.
\end{lemma}

\begin{lemma}
If $\mathit{EXT}$ is executed and the gateway returns a {\tt rate} then {\tt Bob} receives {\tt rate * n} dollars.\\[.5ex]
\end{lemma}

\noindent
\textbf{III. Zero-coupon bond.}
An interesting Findel contract is $ZCB$ (Section~\ref{sec:syntax}): {\tt Alice} should receive from {\tt Bob} 10 dollars, while {\tt Bob} receives 11 dollars after 1 year. Proving that {\tt Alice} is paid is straightforward:

\begin{lemma}
If $ZCB$ is executed, then {\tt Alice} receives 10 dollars from {\tt Bob}.
\end{lemma}

\noindent
One the other hand, due to the semantics of {\tt At} (in Section~\ref{sec:syntax}, {\tt At} is implemented using {\tt Timebound}), we cannot prove that {\tt Bob} receives 11 dollars from {\tt Alice}! What we can prove is that after 1 year, {\tt Bob} can {\it claim} 11 dollars from {\tt Alice}. 
This is because when {\tt Timebound} is executed, a new contract is issued, having {\tt Bob} as owner. 
If {\tt Bob} fails to demand the execution of the contract in the specified time boundaries then he might not receive his money back. 
So, the only guarantee that {\tt Bob} has is given by this theorem:

\begin{lemma}
If $ZCB$ is executed, then {\tt Bob} can claim 11 dollars from {\tt Alice}.
\end{lemma}

\noindent
If {\tt Bob} demands the execution of the contract generated by {\tt Timebound} then he receives the 11 dollars.

%
%
%
%
%

\section{Conclusions}
\label{sec:conclusions}

Using the formal semantics of Findel in Coq we are able to prove properties of Findel contracts. 
When proofs cannot be completed, they typically reveal problems in contracts that might cause money loses (e.g. \textbf{Zero-coupon bond}) or failed agreements (e.g., \textbf{External gateway rate - currency exchange}). This can be done by simply inspecting the failed proof case, which provides the failing conditions.

The results shown in Section~\ref{sec:proofs} use several meta-properties of our Coq formalisation: (1) the step relation preserves state consistency; (2) event generation preserves states consistency; (3) the ledger is consistent; (4) parties cannot withdraw once they have agreed to upon a contract; (5) the time passing is not affected by contract execution.

Our Coq semantics of Findel is small and consists of 450 (non-empty) lines of code.\footnote{The code is available on Github: \url{https://github.com/andreiarusoaie/findel-semantics-coq/}.} The proof of a theorem has around 15 lines of code, but these proofs use several helper lemmas  and some handy tactics that we provide together with the semantics.\\

\emph{Acknowledgements.}
This work was supported by a grant of the ``Alexandru Ioan Cuza'' University of Ia{\c s}i, within the Research Grants program, Grant UAIC, ctr. no. 6/01-01-2017.

\nocite{*}
\bibliographystyle{eptcs}
\bibliography{generic}
\end{document}